\def\ali{A_{\rm Li}}   
\def\gtrsim{\mathrel{\hbox{\rlap{\hbox{\lower4pt\hbox{$\sim$}
}}\hbox{$>$}}}}

\def \SAIT #1 #2 {{\em Mem.\ Soc.\ Astron.\ It.\/} {\bf #1}, #2}
\def \MESS #1 #2 {{\em The Messenger\/} {\bf #1}, #2}
\def \ASTRNACH #1 #2 {{\em Astron. Nach.\/} {\bf #1}, #2}
\def \AAP #1 #2 {{\em Astron. Astrophys.\/} {\bf #1}, #2}
\def \AAL #1 #2 {{\em Astron. Astrophys. Lett.\/} {\bf #1}, L#2}
\def \AAR #1 #2 {{\em Astron. Astrophys. Rev.\/} {\bf #1}, #2}
\def \AAS #1 #2 {{\em Astron. Astrophys. Suppl. Ser.\/} {\bf #1}, #2}
\def \AJ #1 #2 {{\em Astron. J.\/} {\bf #1}, #2}
\def \ANNREV #1 #2 {{\em Ann. Rev. Astron. Astrophys.\/} {\bf #1}, #2}
\def \APJ #1 #2 {{\em Astrophys. J.\/} {\bf #1}, #2}
\def \APJL #1 #2 {{\em Astrophys. J. Lett.\/} {\bf #1}, L#2}
\def \APJS #1 #2 {{\em Astrophys. J. Suppl.\/} {\bf #1}, #2}
\def \APSS #1 #2 {{\em Astrophys. Space Sci.\/} {\bf #1}, #2}
\def \ASR #1 #2 {{\em Adv. Space Res.\/} {\bf #1}, #2}
\def \BAIC #1 #2 {{\em Bull. Astron. Inst. Czechosl.\/} {\bf #1}, #2}
\def \JSQRT #1 #2 {{\em J. Quant. Spectrosc. Radiat. Transfer\/} {\bf #1}, #2}
\def \MN #1 #2 {{\em Mon. Not. R. Astr. Soc.\/} {\bf #1}, #2}
\def \MEM #1 #2 {{\em Mem. R. Astr. Soc.\/} {\bf #1}, #2}
\def \PLR #1 #2 {{\em Phys. Lett. Rev.\/} {\bf #1}, #2}
\def \PASJ #1 #2 {{\em Publ. Astron. Soc. Japan\/} {\bf #1}, #2}
\def \PASP #1 #2 {{\em Publ. Astr. Soc. Pacific\/} {\bf #1}, #2}
\def \NAT #1 #2 {{\em Nature\/} {\bf #1}, #2}

\documentstyle[twoside]{memsait}

\begin{opening}
\title{A non-LTE study of Li~I lines in AGB stars}
\author{Dan Kiselman$^1$, Bertrand Plez$^2$}
\institute{$^1$NORDITA, Blegdamsvej 17, DK-2100 Copenhagen, Denmark. (Current
address: Stockholm Observatory, S-133~36~~Saltsj\"obaden, Sweden.
dan@astro.su.se)\\
$^2$Niels Bohr Institute, Blegdamsvej 17, DK-2100 Copenhagen, Denmark.
plez@nbivax.nbi.dk}
\date{} 
\end{opening}

\begin{document}

\oddpagefooter{}{}{} 
\evenpagefooter{}{}{} 
\
\bigskip

\begin{abstract}
We study the formation of Li lines in an oxygen-rich asymptotic giant branch
star.  
\end{abstract}

\section{Introduction}
Asymptotic giant branch (AGB) stars may be important providers of $^7$Li
to the interstellar medium. It is of importance to determine if
the Li abundance in AGB stars is at the highest levels reported
($A_{\rm Li} \approx 3.0-5.0$) and
if super Li-rich AGB stars do indeed exist.
Abundance analysis of these stars is difficult (see e.g. Plez, Smith
and Lambert 1993) due to the heavy molecular blanketing, uncertainties
in the model atmospheres and possible departures from LTE in the
Li atomic level populations. We address here the latter problem
and somewhat also that of background molecular line formation through
a case study of HV 1963, an S-type AGB star in the SMC (model parameters:
$T_{\rm eff} = 3300~{\rm K}, \log g = 0.0$ [cgs], [Fe/H]$=-0.5$, C/O$=0.2$).
\begin{figure}
\vspace{4.5cm}   
\caption[h]{Observed spectrum of HV1963 at 6707 \AA\ and 8126 \AA\
(dots) together with LTE (dashed line) and non-LTE (full line)
spectra computed in the {\em s0s0} case
for $\ali =  $ 2.0, 3.0, 4.0, 5.0, and 6.0. The spectra were obtained
with the CTIO 4m telescope and have a resolution of about 20000.}
\end{figure}

\section{Methods}
We use POSMARCS plane-parallel, static, LTE model atmospheres with
an opacity sampling treatment of up-to-date opacities (based on
Plez, Brett and Nordlund 1992).

The MULTI 2.0 program (Carlsson 1986, 1991) is used to solve the
non-LTE problem.
We use the 21-level model of Li I
that was compiled by Carlsson et al. (1994) and kindly provided
by Dr. M. Carlsson.
The many molecules and background opacity sources of importance
at these low temperatures cannot be handled by MULTI in the present
setup. Thus, background opacities are provided to MULTI as input files by
the TurboSpectrum package (Plez et al. 1993) for detailed opacities
around the 6104, 6707, and 8126 \AA\ Li I lines, and by POSMARCS for the
other transitions.

\section{Background molecular lines}
The spectra of AGB stars are very crowded with molecular lines (mostly TiO).
We estimate the importance of the assumption (always made in practical
analyses) that molecular lines are formed in LTE, in pure absorption.
Hinkle and Lambert (1975) argue that scattering is more probably the
appropriate mechanism for the formation of molecular lines belonging
to electronic transitions.
To this end, we have done our computations in three different cases:
{\em s0s0} is the standard case (all background line opacity in pure
absorption);
in {\em s0s9} 90 \% of the opacity due to TiO is incorporated in the scattering
coefficient in the line transfer calculation; in {\em s9s9}  the same is also
done when computing the model atmosphere.

\section{Departures from LTE}
Figure 2 illustrates the departures from LTE in the HV1963 model
atmosphere ({\em s0s0}) for $\ali = 3.00$.
The departure coefficients for the level populations
are defined as $b_i = n^{NLTE}_i/n^{LTE}_i$. The main feature in these is the
outward drop which is due to the overionisation resulting from $J > B$
in the photoionisation continua. $b<1$ produces a line that is weaker than
in LTE via a smaller line opacity.

The principal feature for the line source functions ($S_l$) is
the drop in $S_l/B$ caused by photon losses, the effects of which are
propagated down to $\tau > 1$ by scattering in the lines. $S_l/B < 1$ leads
to a strengthening of an absorption line relative to LTE.

The line strength of the 6707 \AA\ line is similar in LTE and in non-LTE
since the overionisation and the depression of the line source function
roughly balance each other.


\begin{figure}
\vspace{7.5cm}   
\caption[h]{Departures from LTE in HV1963 model atmosphere.
Horizontal bars indicate
formation regions of the 6707 \AA\ and the 8126 \AA\ lines by
showing optical depths (0.01, 0.1, 1.0, 10.0) in line cores and continua.
$b_0$ and $b_1$ are the population departure coefficients for the lower
levels of the 6707 \AA\ and the 8126 \AA\ line, respectively. Symbols
show the departure of the line source functions from the local blackbody
value.
}
\end{figure}
\section{NLTE abundance corrections}
Non-LTE abundance corrections are interpolated from theoretical non-LTE
and LTE curves of growth. When the lines of interest are blended, as in our
case, the resulting corrections depend on how the equivalent
widths used for the curves of growth are defined.

The three curves of Fig. 3 show abundance corrections computed
from the numerical equivalent widths.
It is obvious that the effects on Li line formation and atmospheric structure
resulting from the introduction of scattering in the background opacitites
affect the magnitude and sign of the non-LTE corrections for
the 6707 \AA\ line quite significantly.
For the 8126 \AA\ line, corrections are similar for all three cases. The
TiO opacity is less important at that wavelength than at 6707 \AA. Furthermore,
this line is formed at depths where the perturbation in atmospheric structure
due to the change in opacities in {\em s9s9} is small.
\begin{figure}
\vspace{5cm}
\caption[h]{Non-LTE abundance corrections for our standard model with
background
opacities in pure absorption (s0s0), and for the two experiments
with the TiO opacity treated partly as scattering (s0s9 and s9s9). The
corrections are plotted against the abundances an LTE analysis would give
 in each case.}
\end{figure}

By fitting synthetic spectra computed under the different assumptions,
we have estimated the lithium abundance of HV1963.
The result from the 8126 \AA\ line is again rather insensitive to the treatment
of the background molecular opacities.
The {\em s0s0} model gives equal abundances from the two lines:
$\ali \approx 3$.
The models with TiO scattering give inconsistent abundances under the
assumed fundamental parameters.
We do {\em not}, however, consider the scattering
hypothesis ruled out by this limited investigation.

\section{Conclusions}

Non-LTE effects will not rule out the possibility that oxygen-rich
super-Li-rich stars exist. However, the 6707 \AA\ resonance doublet increases
dramatically in strength in atmospheres with a C/O ratio approaching
1 (Smith et al., in preparation). It is thus possible that some
AGB stars show a very strong 6707 \AA\ Li I line but have a "normal"
Li abundance.

The 6707 and 8126 \AA\ lines are not formed in LTE. The magnitude
of the non-LTE corrections is however comparable to other possible
sources of systematic errors: sphericity (up to about
0.3 dex, Plez et al. 1993),
circumstellar absorption, pulsations and shocks.

The 8126 \AA\ line is less sensitive to the detailed treatment
of line blanketing, and presumably insensitive to circumstellar
absorption. It is weaker than the 6707 \AA\ line and less affected by details
of the physics in the outer atmosphere. We therefore recommend observers
to include that line in their programmes. It is however only detectable
for $\ali \gtrsim 2$ in cool AGB stars.

When the Li abundance approaches $\ali = 6$, Li is no longer a trace element.
It becomes an important electron donor, affecting the atmospheric structure.
Atmospheric models should be calculated with the proper Li abundance when
analysing stars with (apparently) very high Li content.

\acknowledgements
B.P. is supported by the Commission of the European Communities.


\end{document}